Advances in the Biomedical Applications of the EELA Project

Vicente Hernández[a], Ignacio Blanquer[a], Gabriel Aparicio[a], Raúl Isea[b], Juan Luis Chaves[b], Álvaro Hernández[b], Henry Ricardo Mora[c], Manuel Fernández[c], Alicia Acero[d], Esther Montes[d] and Rafael Mayo[d]
(On behalf of the EELA Project)

[a] *Universidad Politécnica de Valencia, ITACA – GRyCAP, Spain*
[b] Anteriormente en: *Universidad de los Andes. Actualmente en IDEA, Venezuela*
[c] *Cubaenergía, Cuba*
[d] *CIEMAT, Spain*

**Abstract.** In the last years an increasing demand for Grid Infrastructures has resulted in several international collaborations. This is the case of the EELA Project, which has brought together collaborating groups of Latin America and Europe. One year ago we presented this e-infrastructure used, among others, by the Biomedical groups for the studies of oncological analysis, neglected diseases, sequence alignments and computation phylogenetics. After this period, the achieved advances are summarised in this paper.

**1. Introduction**

Funded by the European Commission, the EELA Project (E-Infrastructure shared between Europe and Latin America) [1] began in January 2006 to build a digital bridge between the existing e-Infrastructure initiatives in Europe and those that were emerging in Latin America, throughout the creation of a collaborative network that shares an interoperable Grid infrastructure to support the development and test of advanced applications.

One of the identified areas of work in scientific research is Biomedicine and, consequently, the applications that can be ported to the Grid. Some of them, falling in the three typical categories of Bioinformatics Applications, Computational Biochemical Processes and Biomedical Models, were selected [2] and started to be deployed on the pilot EELA infrastructures for both production and dissemination purposes.

This document describes the achieved advances in these biomedicine applications in the last year. For a more detailed explanation of them, the reader can consult [2] and [3].

**2. GATE**

The C++ platform based on the Monte Carlo GEANT4 software [4] GATE (GEANT4 Application for Tomographic Emission) [5] models nuclear medicine applications, such as PET and SPECT within the OpenGATE collaboration [6]. The computational-intensive Monte Carlo simulations prevent hospitals and clinical centres from using them for daily practice in radiotherapy and brachytherapy treatment planning. As a result, the objective of GATE is to use the Grid environment to reduce the computing time of Monte Carlo simulations providing a higher accuracy in a reasonable period of time

Nine Cuban centres are currently testing, as users, the results of the simulation of radiotherapy treatments using realistic models that GATE provides in two main oncological problems:
- Thyroid Cancer (the diseases of thyroids are one of the 5 main causes of Endocrinology treatments) [7]; and,
- Treatment of metastasis with $P^{32}$ isotopes [8] by means of the brachytherapy improving the knowledge on the doses captured from the different tissues by accurate simulation.

The EELA consortium has provided the support for the installation and integration of a small Grid site in Cuba in order to run the simplest jobs concerning Monte Carlo simulations. For those cases where the computer load is greater, a previous preparation of the job is done in Cuba and, after that, the necessary material is brought physically to another EELA sites with a higher bandwidth in order to do the submission process.

Besides of that, some researchers have stayed as fellowships for a long period at the ITACA–GRyCAP department (Valencia, Spain) with the support of the EELA funds improving their knowledge in Grid and GATE technologies.

As a conclusion, Grid will increase the performance to the application, but in this case, it will even be more important, since it is an enabling technology opening the doors to a new range of applications and possibilities. All the centres from Cuba currently testing this application bring to the EELA community around 90 cases per month.

## 3. WISDOM

The objective of WISDOM (Wide in Silico Docking of Malaria) is the proposition of new inhibitors for a family of proteins produced *by Plasmodium Falciparum* due to this protozoan parasite causes malaria and affects around 300 million people and more than 4 thousand people die daily in the world. The cross-resistance to antimalarials produced by the focus on a limited number of biological targets has derived in a drug resistance for all classes of antimalarials except artemisinins. Because of this the development of new drugs with new targets is necessary, but the process is cost so the economic profit is not clear for the drug manufacturers.

This WISDOM application consists on the deployment of a high throughput virtual screening platform in the perspective of *in silico* drug discovery for neglected diseases.

The interest of the EELA partners centred in three actions:
- The study of new targets for new parasitory diseases.
- The selection of new targets for malaria.
- The contribution with resources for the WISDOM Data Challenge.

The WISDOM platform [9] has performed in the Fall of 2006 its second High-Throughput virtual Docking of million of chemical compounds available in the databases of ligands to several targets. In this WISDOM Data Challenge-II [10] ULA has proposed two targets in *Plasmodium Vivax* which have been accepted by the consortium and docked within the

EELA as was agreed by the WISDOM DC-II decision-makers. This process has been coordinated by UPV and all the EELA sites have acted as donor of computational and storage resources. The results will be presented in future works.

## 4. BLAST in Grid

The study of the functionality of the different genes and regions is one of the most important efforts on the analysis of the genome. If the queries and the alignments are well designed both functional and evolutionary information can be inferred from sequence alignments since they provide a powerful way to compare novel sequences with previously characterised genes.

The Basic Local Alignment Search Tool (BLAST) [11] finds regions of local similarity between sequences. The program compares nucleotide or protein sequences to databases and calculates the statistical significance of matches. This process of finding homologous of sequences is very computationally-intensive since the searching alignment of a single sequence is not a costly task, but normally, thousands of sequences are searched at the same time.

The biocomputing community usually relies on either local installations or public servers, such as the NCBI [12] or the gPS@ [13], but the limitations on the number of simultaneous queries make this environment inefficient for large tests. Moreover, since the databases are periodically updated, it will be convenient to do the same with the results of previous studies.

BLAST has been ported to the Grid according to different approaches [14] because the number of fragments to be analysed and the periodical updating of the information will be increased. EELA has adopted the availability of an independent Grid-enabled version integrated on the Bioinformatics Portal of the Universidad de los Andes [15] providing registered users with results in a shorter time. Basically the users access the service through this web portal and, after that, to the EELA Grid with a Gate-to-Grid, i. e., an EELA Grid node which provides a WSRF-Based web interface. The security is provided by means of a MyProxy server which generates manually and temporally certificates that will be retrieved by the UI when required.

BLAST in Grid (BiG) has been used for searching similar sequences and inferring their function in parasite diseases such as the Leishmaniasis (mainly *Mexican Leishmania*), Chagas (mainly *Trypanosoma Cruzi*) and Malaria (mainly *Plasmodium vivax*).

## 5. Phylogeny (MrBayes)

A phylogeny is a reconstruction of the evolutionary history of a group of organisms used throughout the life sciences, as it offers a structure around which to organize the knowledge and data accumulated by researchers. Computational phylogenetics has been a rich area for algorithm design over the last 15 years. The inference of phylogenies with computational methods is widely used in medical and biological research and has many important applications, such as gene function prediction, drug discovery and conservation biology [16].

The most commonly used methods to infer phylogenies include cladistics, phenetics, maximum likelihood, and Markov Chain Monte Carlo (MCMC) based Bayesian inference. These last two depend upon a mathematical model describing the evolution of characters observed in the included species, and are usually used for molecular phylogeny where the characters are aligned nucleotide or amino acid sequences.

Due to the nature of Bayesian inference, the simulation can be prone to entrapment in local optima. To overcome local optima and achieve better estimation, the MrBayes program [17] has to run for millions of iterations (generations) which require a large amount of computation time. For multiple sessions with different models or parameters, it will take a very long time before the results can be analyzed and summarized.

The phylogenetic tools are widely demanded by the Latin America bioinformatics community. A Grid service for the parallelised version of MrBayes application is currently being developed and a simple interface will be deployed on the bioinformatics portal of Universidad de los Andes [18] in a similar way to that done for BiG. Nevertheless, some previous works with this parallelization have been successfully submitted and executed in the EELA infrastructure and been demonstrated in the first EELA conference [19].

### 6. Conclusion and new challenges

The EELA e-infrastructure is permitting various collaborative groups in Latin America to use more powerful computational resources than those available on their centers. This achieves that the lines of investigation stated in the document can be feasible as the computational requirements for them are being met and new results are being offered to the scientific community.

The EELA project currently has four biomedical pilot applications running, but is also looking for new ones. For this purpose, it is offering its support with initiatives like Grid School [20]. As a result, new biomedical applications such as EMBOSS [21] will be ported to the Grid very soon. With this free Open Source software analysis package specially developed for the needs of the molecular biology user community EELA will keep on working in the Biomedical field.

### References

[1] http://www.eu-eela.org
[2] EELA Deliverable D3.1.1. http://www.eu-eela.org/eela_deliverables_list.php
[3] M Cárdenas, V Hernández, R Mayo, I Blanquer, J Pérez-Griffo, R Isea, L Nuñez, HR Mora, M Fernández, "*Biomedical Aplications in EELA*", Studies in Health Technology and Informatics 2006;120;397-400.
[4] http://geant4.web.cern.ch/geant4/
[5] S. Jan, et al., *"GATE: a simulation toolkit for PET and SPECT"*, Phys. Med. Biol. 2004;49;4543-4561

[6] http://opengatecollaboration.healthgrid.org/

[7] D. Navarro *"Epidemiología de las enfermedades del tiroides en Cuba"*, Rev Cubana Endocrinol. 2004;15.

[8] J. Alert, J. Jiménez, *"Tendencias del tratamiento radiante en los tumores del sistema nervioso central"*, Rev Cubana Med 2004; 43(2-3).

[9] http://wisdom.healthgrid.org/

[10] http://wisdom.healthgrid.org/index.php?id=139

[11] http://www.ncbi.nlm.nih.gov/Education/BLASTinfo/information3.html

[12] http://www.ncbi.nlm.nih.gov/

[13] http://gpsa.ibcp.fr/

[14] G Aparicio, S Götz, A Conesa, JD Segrelles, I Blanquer, JM García, V Hernández, "*Blast2GO goes Grid: Developing a Grid-Enabled Prototype for Functional Genomics Analysis*", Studies in Health Technology and Informatics 2006;120:194-204

[15] http://www.cecalc.ula.ve/blast/

[16] K. Lesheng, *"Phylogenetic Inference Using Parallel Version of MrBayes"*. http://www.nus.edu.sg/comcen/svu/publications/hpc_nus/sep_2005/ParallelMrBayes.pdf

[17] http://mrbayes.csit.fsu.edu/

[18] http://portal-bio.ula.ve/

[19] http://documents.eu-eela.org/getfile.py?recid=273

[20] http://www.eu-eela.org/egris1/index.html

[21] http://emboss.sourceforge.net/what/#Overview